# DESIGN AND IMPLEMENTATION OF A MEASUREMENT-BASED POLICY-DRIVEN RESOURCE MANAGEMENT FRAMEWORK FOR CONVERGED NETWORKS

**S. Y. Yerima[1], G.P. Parr[2], S. McCLean[3], P. J. Morrow[4] and K. Sivalingam[5]**

*India – UK Advanced Technology Centre (IU–ATC) of Excellence in Next Generation Networks Systems and Services*

[1,2,3,4]*School of Computing and Information Engineering, University of Ulster, Northern Ireland*

E-mail: [1]s.yerima@ulster.ac.uk, [2]gp.parr@ulster.ac.uk, [3]si.mcclean@ulster.ac.uk, [4]pj.morrow@ulster.ac.uk

[5]*Indian Institute of Technology Madras, India*

E-mail: krishna.sivalingam@gmail.com

***Abstract***

***This paper presents the design and implementation of a measurement-based QoS and resource management framework, CNQF (Converged Networks' QoS Management Framework). CNQF is designed to provide unified, scalable QoS control and resource management through the use of a policy-based network management paradigm. It achieves this via distributed functional entities that are deployed to co-ordinate the resources of the transport network through centralized policy-driven decisions supported by measurement-based control architecture. We present the CNQF architecture, implementation of the prototype and validation of various inbuilt QoS control mechanisms using real traffic flows on a Linux-based experimental test bed.***

## 1. INTRODUCTION

Efficient control and management infrastructure are needed to provide coordinated, scalable and transparent resource management and QoS control as fixed and wireless networks converge towards IP-based transport in next generation networks. In order to meet this requirement however, the complexity of configuration, control and management operations needed to support transparent service provisioning must be overcome. Policy-based network management (PBNM) is one promising approach that provides this capability by easing the management of complex networks *through automated and distributed* structures using centralized policies. To this end, we have developed a QoS management framework, CNQF (Converged Networks QoS Management Framework) based on the PBNM paradigm.

CNQF is aimed at providing homogenous, unified and adaptive measurement-based QoS control and resource management over heterogeneous access technologies. By leveraging PBNM paradigm, the CNQF architecture provides the means for application transparency across existing and emerging access technologies, thus permitting applications to be transport-layer agnostic when deployed.

As part of the ongoing IU-ATC (India-UK Advanced Technology Centre of excellence) project, an experimental CNQF framework prototype is being built to provide a platform for development and evaluation of advanced algorithms and mechanisms for policy-based QoS management in converged next generation networks. The CNQF prototype is being developed in Java within a configurable testbed designed to provide representative converged networks scenarios for tests and evaluations. This paper presents the implementation of CNQF subsystems and entities as well as initial experiments conducted to test and validate various underlying mechanisms enabling QoS control and management within the CNQF architecture.

The rest of the paper is organized as follows. The next section provides the background and motivation for our work. Section 3 explains the CNQF architecture design and the constituent subsystems. Section 4 deals with the implementation of the CNQF prototype as well as the evaluation test bed configuration. Section 5 presents the tests conducted on the testbed to validate the operation of the current CNQF implementation.

## 2. BACKGROUND AND MOTIVATION

One of the key advantages of Policy-based network management (PBNM) is that it can simplify the administration of complex operational characteristics of a network, including QoS, access control, network security, and IP address allocation [1]. The PBNM architectures published by the various standardization bodies can be found in [2], [3], and [4], for example. Telecoms and Internet Converged Services and Protocols for Advanced Networks (TISPAN) technical committee of the European Telecommunications Standards Institute (ETSI) has defined a Resource and Admission Control Subsystem (RACS) consisting of a Service-based Policy Decision Function (SPDF) and Access Resource Admission Control Function (A-RACF) [3]. Both of these interact with Policy Enforcement Points (PEPs) in the underlying networks. In that regard, the architecture shares similarity with the IETF policy model which specified a policy enforcement point (PEP) and a Policy Decision Point (PDP) as part of its architecture [4]. Similarly, Third Generation Partnership Project (3GPP) defined a Policy Decision Function (PDF) in their Release 5/6 policy framework [2].

While the standards bodies have defined architectures, protocols and interfaces that are crucial to interoperability of disparate vendor equipments that conform to the same standards, details of implementation are left out and are usually vendor-specific. Hence, with new wired and wireless technologies emerging coupled with the need to address their convergence and management requirements, PBNM based framework and architectures are still being actively researched.

Kim et. al., for example, present an IP QoS management framework in [5] designed to provide QoS control in ad hoc military environments. The framework is based on SNMP and DiffServ. Similarly, a policy-based multi-layer QoS architecture for network resource control based on policy-based routing and Traffic Engineering (TE) is presented in [6]. In [7] Oziany et. al. present an XML-driven QoS management framework for IMS based networks. Other works such as [1], [8]-[14] can be found employing PBNM in different contexts including VPNs, Multi-hop ad hoc networks, MPLS-enabled networks and virtualization environments.

An important distinguishing feature of the CNQF design presented in this paper from the aforementioned is the incorporation of context management functionality as an important building block of its architecture thus enabling added intelligence to provide adaptive policy-driven decisions within the framework. Furthermore, our work also contributes in bridging the gap between the PBNM architectural definitions within the standards on the one hand, and the implementation and experimentation on the other hand which serves to provide useful insight gained through evaluation studies. Hence, this paper not only presents the CNQF architectural design framework but also implementation of key subsystems and experimental studies to validate their operation.

## 3. THE CNQF ARCHITECTURE

The CNQF architecture is presented in this section. It is designed with three *logical* subsystems including: Measurement and Monitoring Subsystem (MMS), RMS (Resource Management Subsystem) and the Context Management and Adaptation Subsystem (CAS). While functional elements within the subsystems are designed to be integrated horizontally, they also form part of a hierarchical structure as is typical of PBNM systems [7]. The hierarchical structure is depicted in Fig.1.

The top level consists of tools that provide centralized administrative capabilities such as Graphical User Interfaces (GUI) for high level configuration, and policy entry/editing; visualization tools for network-wide configuration and status monitoring; and central high-level repositories. The PDL (Policy Decision Layer) comprises of various Policy Decision Points (PDPs) such as the Resource Brokers that form part of the RMS; these are centrally managed via the PAL (Policy Administrative Layer). The bottom layer is the MCAL (Measurement, Control and Adaptation) layer that consists of elements that directly interface with the Policy Enforcement Points (PEPs) in the transport network. These elements include the various Resource Controllers that are part of the RMS and also the Network Monitors that are part of the Measurement and Monitoring Subsystem.

### 3.1 RESOURCE MANAGEMENT SUBSYSTEM (RMS)

The RMS within CNQF framework is primarily designed to provide co-ordination, control, and allocation of resources along the end-to-end transport path of the CNQF QoS domain. The subsystem is structured such that the underlying allocation/control mechanisms could be based on simple static policies to complex, dynamic policies driven by measurement-based resource control algorithms. These mechanisms are present within the Resource Brokers, which are responsible for the policy-based decision within the RMS. In a CNQF QoS domain with wireless/fixed access edge networks as well as a core network, the RBs could be Wireless Access Resource Brokers (RBs), Fixed Access Resource Brokers (FARB) or Core Network Resource Brokers (CNRB).

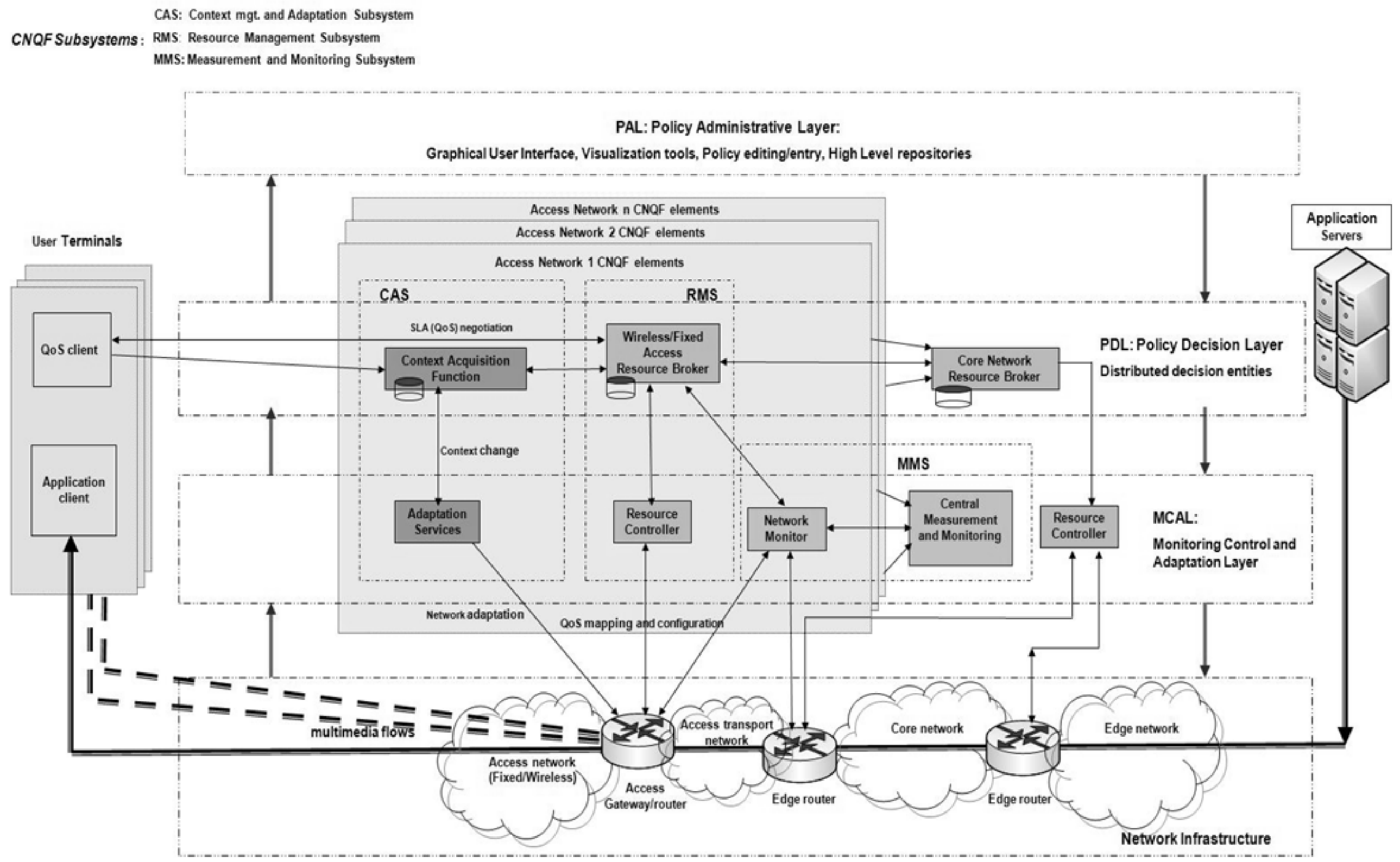

Fig.1. CNQF architecture

In such configuration, the WARB and FARB will interface with the CNRB as shown in Fig. 1. This then allows for inter FARB/WARB resource brokerage by the CNRB thus enabling scalable end-to-end management of resources along the transport plane within the QoS administrative domain.

Another entity within the RMS is the lower layer MCAL element, the Resource Controller (RC). The RCs implement the logic of the policy actions that enable (re)configuration of QoS mechanisms within the PEPs such as routers, gateways, switches and other key nodes within the transport plane where QoS mechanisms are implemented. Each set of policy actions enabled by the RC is mapped to specific policy condition(s) evaluated within decision entities (RB) in response to events defined within the QoS policies.

Each RB (WARB, FARB, CNRB) is interfaced with one or more corresponding RCs (FARC, WARC, CNRC) in the MCAL layer. The RCs perform different configuration and control functions depending on where the PEPs are located on the transport plane. For instance, an RC located in the edge router (PEP) may be responsible for packet marking (e.g. DiffServ Code Points, DSCP marking in a DiffServ domain) in response to CNRB policy decisions. While in an edge wireless access network, an RC may be responsible for configuration of gateway nodes to dynamically map layer 2 QoS parameters (e.g. WiMAX QoS classes) to layer 3 IP QoS parameters (e.g. DiffServ DSCPs or MPLS LSPs) for different flows.

### 3.2 MEASUREMENT AND MONITORING SUBSYSTEM (MMS)

The ability to monitor all network devices and network elements is vitally important to the operation of CNQF. The MMS provides this capability within the framework by incorporating both passive and active measurement capabilities. MMS facilitates *closed-looped, adaptive and measurement-based QoS control* without which CNQF will be limited to providing open-loop QoS provisioning based on, for example pre-determined end-to-end resource allocation derived from a priori QoS negotiations. With MMS in the loop, fine-grained resource allocation and QoS control could be achieved through feedback of measurement data to the RMS.

The MMS consists of distributed network monitoring entities (NMs) located at the PEPs for measurement and monitoring collection. These NMs also form part of the MCAL layer. Centralized measurement capability is provided by another MMS entity, the CMM (Central Measurement and Monitoring). If required, the NMs interface to the CMM which serves as an aggregating entity for the entire MMS and could the provide high level summaries that are useful for gauging the health of the network via visualization interfaces on a centralized management station for example. The NMs passive and active measurement mechanisms are explained in section 4 where their implementation within a Java-based CNQF prototype is discussed.

### 3.3 CONTEXT MANAGEMENT AND ADAPTATION SUBSYSTEM (CAS)

PBNM systems such as CNQF stand to benefit from the use of *context information* to drive policies/policy adaptation. This is because context information equips the management system with increased intelligence and ability to adapt service provision, resource allocation, and QoS control in a more flexible and efficient manner. It also gives more autonomy to the system to respond to highly dynamic operational conditions. For example, resource allocation may be made responsive to different user contexts such as location, time, device capability, battery capacity etc. Through *context-awareness*, the PBNM system may apply different resource management policies to different 'contexts'. For example a user may receive different bandwidth allocations or may be re-assigned to a different QoS class in different locations if the network is aware of the user's location (context) and is able to allocate location-dependent usage through context-aware policies.

As shown in Fig.1, CNQF provides context-aware functionality through its Context Management and Adaptation Subsystem (CAS). CAS consists of distributed Context Acquisition Function blocks (CAF) instantiated in each access network. The CAFs are PDPs that execute context-aware or context-driven policies within the CNQF system. Each CAF elements has associated Adaptation Servers (ADs) which are function blocks that configure/reconfigure PEPs directly affected by context-driven policy decisions in the CAF. Entities that can be characterised by context within the CNQF PBNM system could be physical objects e.g. a user device, router, switch, gateway node, physical link, wireless channel; or could be a virtual object such as MPLS path, or a VPN tunnel.

An example use case scenario involving CAS within a CNQF administered wireless network is as follows. The CAF entity would direct the ADs element to configure the Radio Access Network (RAN) node according to predefined context-aware handover management policies, where e.g. a pre-configured user profile, location, or speed of user device provides the 'context information' for executing a particular network-centric handover strategy according to policies. Other exemplary use case scenarios for CNQF based context-driven QoS control and resource allocation can be found in our previous work [15] and [16].

## 4. CNQF FRAMEWORK IMPLEMENTATION

The CNQF architecture which is designed as a PBNM system with distributed entities within a layered structure has been presented in the previous section. In this section, we present the implementation of a Java-based working CNQF prototype built using the open-source *NetBeans* IDE 6.9 platform. Presently, distributed entities of the MMS and RMS subsystems have been implemented providing capability for adaptive, measurement-based QoS control based on the CNQF architecture.

### 4.1 JAVA BASED CNQF PROTOTYPE

#### *4.1.1 RMS Implementation: Resource Broker/Resource Controller Chain:*

As mentioned earlier, the resource management decision logic which are designed to be driven by high-level policies reside within the RBs. The RBs communicate with the RC elements which implement the policy actions at the designated

PEP. The CNQF RB is built such that high-level policies entered within a GUI policy editor in the administrative layer are mapped into a set of *commands* (written in Java) within the decision logic. The implementation of the decision logic will differ depending on whether the RB plays the role of CNRB, WARB or FARB. The commands are Java code segments that invoke the services of instances of other MCAL entities i.e. network monitoring and resource control elements that are installed and running at the PEPs within the network. For instance, a high level CNQF policy that has an action part*: Configure Edge Router* will be mapped to the following Java code segment within the RB:

*new ResCon*

*ResCon.ConfigureEdge ()*

This creates an instance of the Resource Controller *interface* within the RB that in turn calls the remote *ConfigureEdge()* method (which provides remote edge router configuration services for the RB policies) using Java RMI (remote Method Invocation) technology. The corresponding remote RC instances are installed on the PEPs and these are *implemented as ResConImpl* class. *ResConImpl* implements the methods such as *ConfigureEdge()*, *ConfigureCore()*, and a host of others that can be invoked by the RB via the RMI communication interface. Details of the methods implementation will depend on the PEP type i.e. whether it is a router, switch or gateway and also the specific APIs available for interacting with the internal mechanisms. Thus, the *ResConImpl* class is the wrapper class which can be customized to wrap the functionality of specific QoS mechanisms within the heterogeneous PEPs thus exposing a homogenous API and enabling technology independent RMS to be achieved.

Since the current CNQF prototype implementation is deployed on a testbed with Linux-based routers as the key PEPs, the *ReSConImpl* class which implements the RC functionalities within the various methods utilizes Linux Kernel APIs. This allows for interaction with the Linux TC (traffic control) utility and other utilities for configuration of the routers in response to the *policy actions* invoked by the RB decision logic.

#### *4.1.2 MMS Implementation: NetMon Class:*

The network monitoring entity NM of the MMS has been implemented as a *NetMon* class. This allows for closed-loop and autonomous QoS control within the prototype. Both active and passive measurement capabilities have been incorporated. The passive measurement aspect is based on SNMP using available Management Information Blocks (MIBs). Thus, CNQF can create and install instances of NetMon at various PEPs where desired provided they are also SNMP-enabled PEPs. The active measurement functionality is based on probe packets injected into the network to provide measurements of loss and delay metrics.

#### *4.1.3 MMS Passive Measurement Implementation:*

CNQF passive measurement functionality is provided within the NetMon via the SNMP protocol. During runtime/deployment, a centralized NetMon instance can be configured to poll network-wide measurements using the IP addresses of the routers' interfaces for example. This constitutes the Centralized Measurement and Monitoring (CMM) mode.

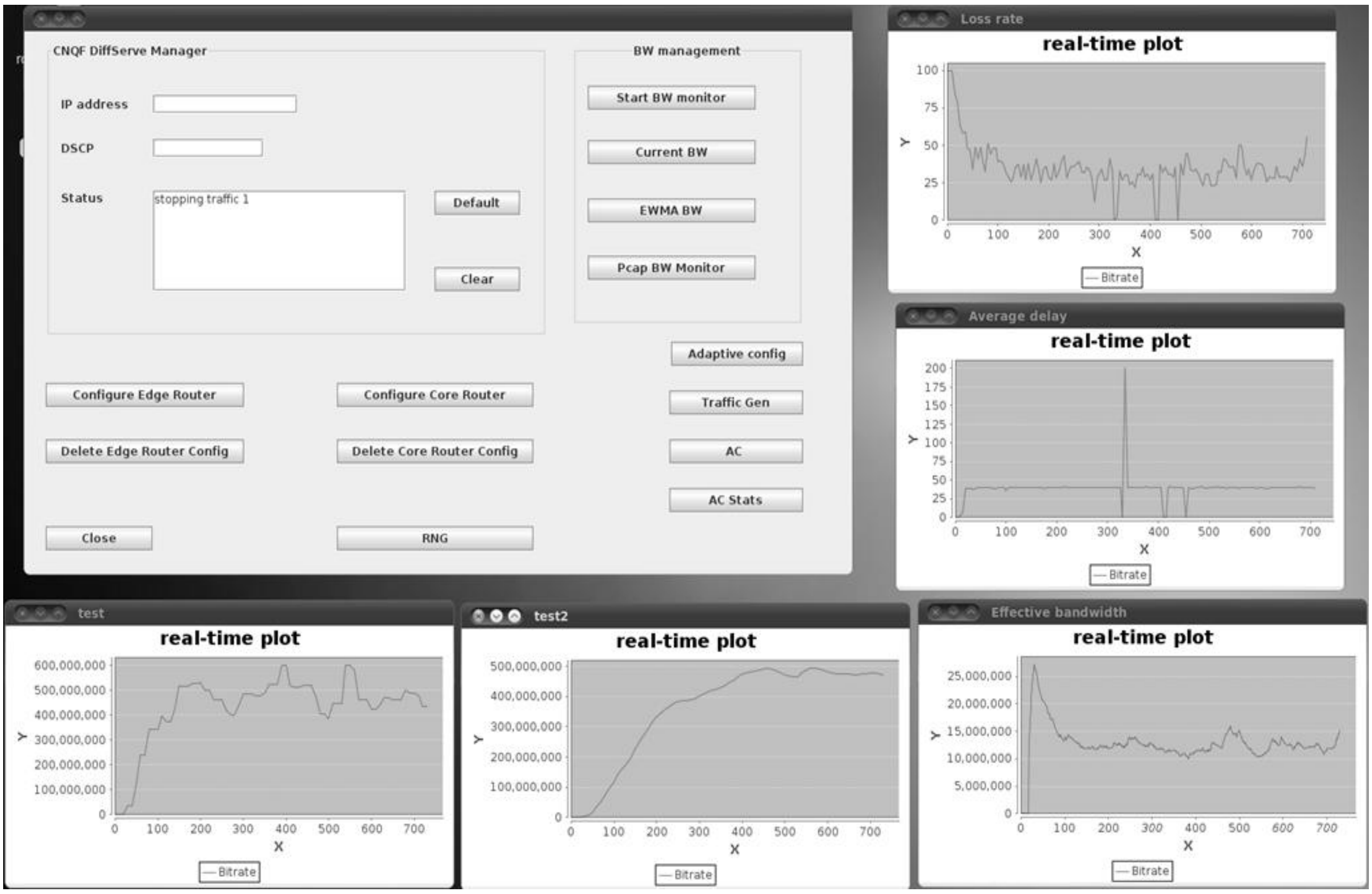

Fig.2. Sample snapshot of an admin interface within the current CNQF prototype

Alternatively the *NetMon* instances can run on each router while pre-processing and sending measurement data on demand to other CNQF entities thereby minimizing control/management network traffic overhead. The CNQF NetMon SNMP agent is built using *SNMP4j* [17], an open source object oriented SNMP API for Java managers and agents. SNMP4j supports command generation (manager mode) and command responding (agent mode) as well as synchronous and asynchronous requests. Table.1 illustrates the RFC 1213 MIB OIDs (Object IDs) used within the *NetMon* class for bandwidth monitoring.

Table.1. MIB OIDs used in CNQF NetMon class (RFC 1213)

| MIB object | Description | OID |
|---|---|---|
| ifInOctets | The total number of octets received on the interface, including framing characters | 1.3.6.1.2.1.2.2.1.10.2 |
| ifOutOctets | The total number of octets transmitted out of the interface, including framing characters | 1.3.6.1.2.1.2.2.1.16.1 |
| ifSpeed | An estimate of the interface's current bandwidth in bits/sec | 1.3.6.1.2.1.2.2.1.5.1 |

From the MIB objects, *NetMon* calculates the interface or link bandwidth using:

$$BW\ (bits/s) = (O(t) - O(t-\Delta t) *8)/ \Delta t \quad (1)$$

Where $\Delta t$ is the interval between two SMNP *get* operations that are used to read the MIB values O(t) which is basically a counter indicating the number of octets sent (ifOutOctets) or received (IfInOctets) on the network interface. Since the MIB variables are stored as counters, two poll cycles are taken by the *NetMon* instance and the difference is calculated to get the bandwidth. Utilization is calculated using:

$$BWU\ (\%) = (O(t) - O(t - \Delta t) *8*100)/ (\Delta t * ifSpeed) \quad (2)$$

Within *NetMon* class, the average bandwidth is also tracked using the exponentially weighted moving average:

$$BW(t) = (1-\alpha) * BW\ (t - \Delta t) + \alpha * BW(t) \quad (3)$$

The QoS policies processed within the RBs would typically leverage these passive measurements collected by the *NetMon* instances to influence policy decisions (for example those related to bandwidth management).

#### *4.1.4 MMS Active Measurement Implementation:*

The *NetMon* active measurement mechanism is implemented using two main techniques: packet capture and network probes. This has been built in order to allow CNQF prototype autonomously derive network metrics that are unavailable from the use of SNMP MIBs.

Packet capture is implemented within *NetMon* with an open-source version of *jNetPcap* [18]. jNetPcap is an open-source Java library that contains a Java wrapper for nearly all *libpcap* library native calls. (*libpcap* is a portable C/C++ library for network traffic capture which allows for 'sniffing' the network from within an application). jNetPcap decodes captured packets in real-time and also provides a large library of network protocols. Furthermore, users can easily add their own protocol definitions using Java SDK. With the packet capture mechanism inbuilt, *NetMon* is able to perform real-time monitoring at any network interface within the CNQF QoS domain.

*NetMon* also uses the open-source Bwping utility to send probe packets into the network. Bwping enables estimation of bandwidth, packet loss and response times between two hosts. It uses ICMP (Internet Control Messaging Protocol) echo request/reply mechanisms and does not require any special software on the remote host, only the ability to respond to ICMP echo request messages.

Policy-based admission control is one aspect where the MMS *NetMon* entities can be leveraged for adaptive, closed-loop PBNM functionality. Real-time measurements of bandwidth, loss and delay are fed into the RB decision engines where admission control decisions are made using algorithms that exploit the *NetMon*-measured QoS metrics. Furthermore, the currently implemented passive and active *NetMon* mechanisms allow the CNQF prototype to build a map of the network state in real-time. Several in-built graph visualization tools for real-time monitoring are also present within the CNQF admin interface some of which are depicted in Fig.2.

## 4.2 LINUX BASED TESTBED IMPLEMENTATION

CNQF prototype is constructed and evaluated on a Linux-based testbed. The configuration is shown in Fig.3. The CNQF testbed consists of two Linux-based edge routers and a Linux-based core router. These elements constitute the PEPs each having an instance of CNQF RC (ResConImpl) that interacts with the Linux router kernel to set various parameters that enable (re)configuration of QoS management strategies stipulated in the high-level policies processed by the RB.

As mentioned earlier, the Linux TC (traffic control) utility in the kernel provides commands for implementing packet marking, classification, queuing disciplines, and policing of flows (enabling transport layer QoS mechanisms). Within the testbed, the RCs employ TC commands for low level configuration which have equivalent mappings to the RB Java code that implement the high level *policy actions*. The testbed elements include:

- *CNQF management station*: houses central CNQF management application with the GUI policy editing tool and RMS CNRB implemented in Java which invokes policy actions via remote RCs (ResConImpl instances) installed at the PEPs (routers).
- *Edge routers A and B*: Ubuntu 10.0.4 Linux PCs with 2.66 GHz Intel Xeon, 3GB RAM, configured as edge routers with TC utility installed to enable configuration of the router interface(s) for ingress packet marking, and for egress classification, queuing and policing via RC's response to CNQF policy decisions.
- *Core router*: Ubuntu 10.0.4 Linux PC with 2.66 GHz Intel Xeon, 3GB RAM, with TC utility installed to enable configuration of packet classifiers and filters through CNQF policies also via an RC instance.

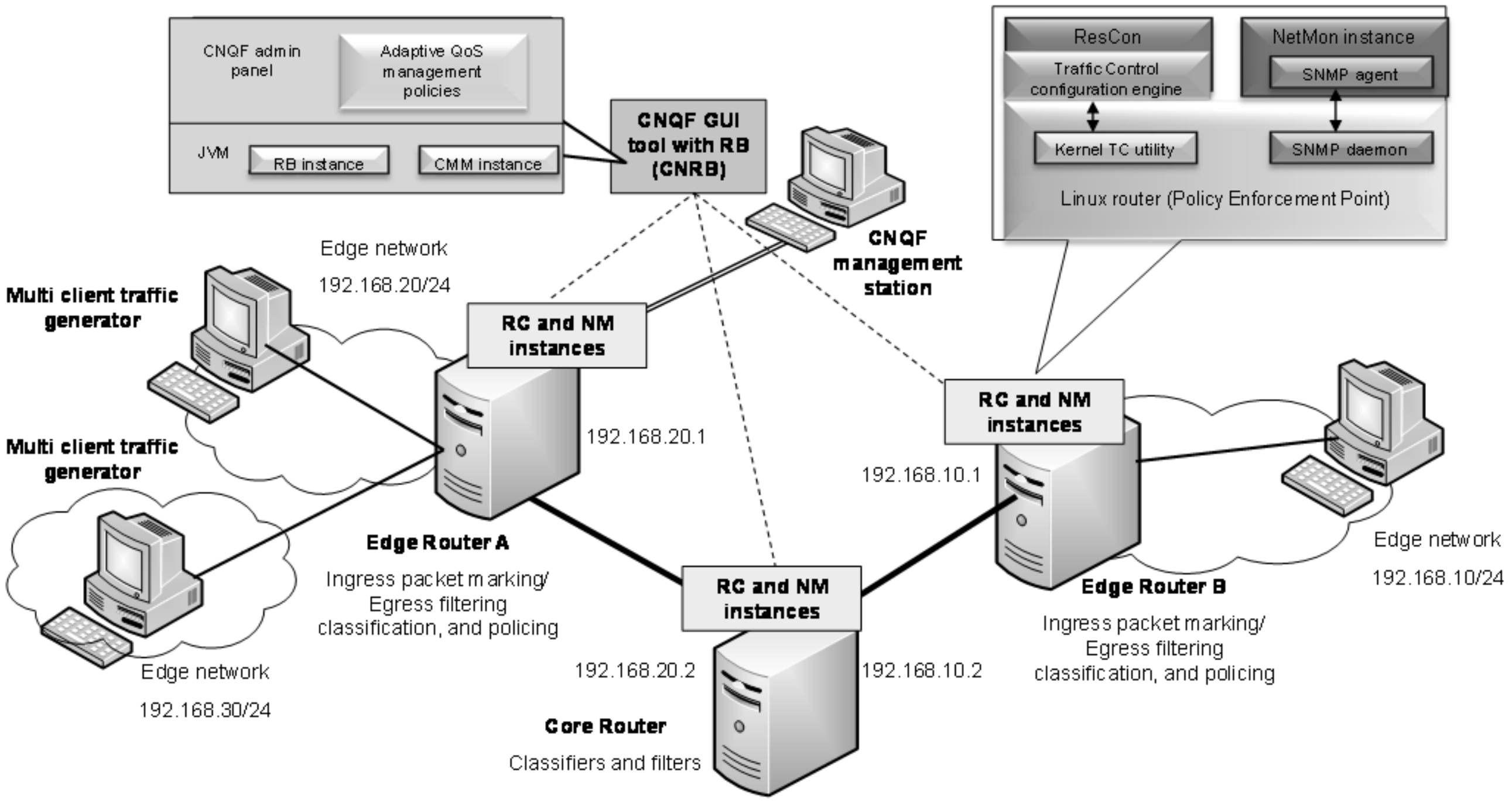

Fig.3. CNQF development and evaluation test bed

- *Traffic generators*: The *Ntools* [19] traffic generator is used in some experiments to generate multi-client traffic with different flow characteristics including constant bit rate (CBR), On-Off traffic, and variable bit rate (VBR) traffic.

# 5. CONDUCTED PROTOTYPE VERIFICATION TESTS

In order to validate key aspects of the prototype we carried out experiments with real and generated traffic flows on the testbed. First the flows will be observed in a baseline scenario where CNQF is not deployed or enabled. Then we will make a comparative analysis with the case where the CNQF functionalities are in place or enabled in order to evaluate CNQF's impact on QoS configuration and control.

The key functionalities we want to validate are the RMS *ResCon* implementation and *NetMon* measurement and monitoring mechanisms. These elements form the core functionality needed to enable closed-loop adaptive QoS control and resource management within the CNQF framework.

## 5.1 TRAFFIC QOS MANAGEMENT CONFIGURATION VIA RESCONIMPL

As described in section 3, the RBs process high-level policies to drive decision making which triggers the actions to be taken in response to policy conditions within the policy rules. Recall that the RC (implemented as ResConImpl class) is the element responsible for configuration of the QoS mechanisms within the PEPs i.e. edge routers and core routers in our testbed. Hence, the RC contains the logic to configure the parameters via the Linux TC API. Note that the same principle can be extended to other kinds of PEPs. This allows for CNQF implementation in large scale operational networks with heterogeneous PEPs existing within the transport plane, since the RCs configuration logic is meant to wrap the functionality of the PEPs whilst hiding the implementation or configuration commands from the (technology-independent) RB high level declarative policies.

In our tests we employ CNQF to configure the network for DiffServ IP QoS management such that traffic coming from edge networks attached to the ingress/egress routers on the testbed may be classified into different DiffServ QoS classes. The steps to achieve this via CNQF are:

1. Specify the high level policy rule for edge routers' configuration to be processed by the handling RB. e.g.:

   Policy rule 1: *If src.ip == 192.168.20.X MARK packets with DSCP ==0x2e*

   which stipulates that all packets from edge network 192.168.20.X/24 will be marked with 0x2e within the DS field of the IP header.
2. A command to configure the edge routers (ingress interfaces) will be issued from the GUI running on the CNQF management station. This will invoke the remote RC instance to configure the edge routers with the TC commands:
   - *tc class change dev eth0 classid 1:1 dsmark mask 0x0 value 0xb8*
   - *tc filter add dev eth0 parent 1:0 protocol ip prio 1 u32 match ip src 192.168.20.10/24 flow id 1:1*
3. Specify the high level policy rule for core routers' configuration to be processed by the handling RB. e.g.:

   Policy rule 2: *If DSCP ==0x2e QUEUE packets with PRIORITY 1*

   which stipulates that all packets with DSCP marked with 0x2e within the DS field of the IP header will be queued with higher priority i.e. to receive Expedited Forwarding (EF) Per-hop-behavior (PHB). Thus packets with DSCP=0x2e will be forwarded with higher priority within the core routers.

4. A command to configure the core routers (egress interfaces) will be issued from the GUI running on the CNQF management station. This will invoke the remote RC instance to configure the core routers with the TC command:

   *tc filter add dev eth0 parent 1:0 protocol ip prio 1 u32 match ip tos 0xb8 0xfc flow id 1:1*

Note that these configurations could be conducted dynamically by specifying adaptive policies that could autonomously change the configurations to suit different contexts or network conditions. But for the purpose of our validation tests, traffic from edge networks connected to the testbed will be observed using the CNQF *NetMon* entities and a comparison will be made between scenarios where the above policies are disabled and where they have been applied according to the steps outlined previously.

## 5.2 VALIDATION TESTS

The policies enabled within the CNQF in the previous section allow packets from a particular edge network to get DiffServ EF priority treatment by tagging the packets with EF DSCP. Expectedly, without these polices enabled, the tagged traffic from that particular edge network will have to compete with untagged flows from the other edge networks for resources. Fig.4 shows the aggregate traffic from all the connected edge networks captured using the *NetMon* passive measurement mechanism built with SNMP4j. The figure depicts the temporal variation in bandwidth (link) utilization on the egress interface of the ingress edge router A (see Fig.3). Note that the total link capacity between edge router A and the core router B is 1Gps.

The flows from the edge networks have been configured to arrive at the ingress edge router A with exponentially distributed inter-arrival times using the open source Ntools traffic generator. After an exponentially distributed period, each arriving flow is terminated. During this time the tagged flow from the edge network is observed at both ingress and egress routers using the installed CNQF *NetMon* instances to determine the effect of configuration policies on the flow. Fig.4 shows that the aggregate traffic at router A peaks at around 600 Mbps, while at around time 810s from the start of the experiment, the last arriving flow has been terminated.

### *5.2.1 Validation Tests Without CNQF Configuration:*

During the first experimental scenario (where the CNQF QoS policies were not applied) the aggregate traffic as observed from Fig.4 was applied to the testbed. At the same time, we observe that 1 Mbps flow from the 192.168.20.x/24 edge network captured at the ingress of the router A depicted in Fig.5 showing a consistent pattern. Simultaneously, the *NetMon* instance at the egress edge router where the link capacity (between router C and router B) is 100 Mbps shows considerable degradation of the flow QoS as a result of congestion within the edge-to-edge link from router A to B (Fig.6). The congestion phenomenon occurs when the aggregate traffic (Fig.4) leaving the egress interface of the ingress edge router A approaches and exceeds the bottleneck edge-to-edge capacity of 100 Mbps.

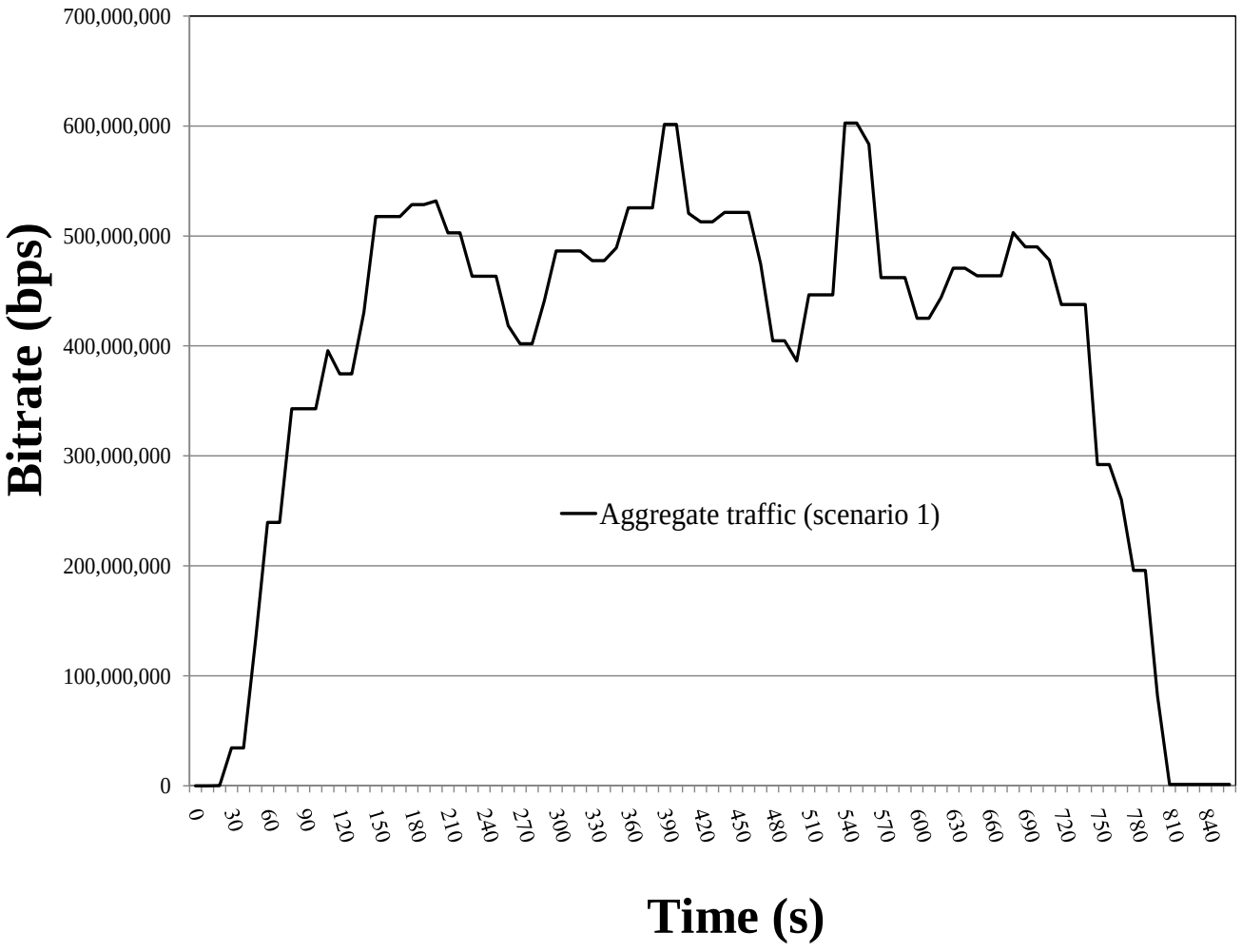

Fig.4. Aggregate traffic from all edge networks to the test bed ingress router observed from the CNQF NetMon tool (scenario1)

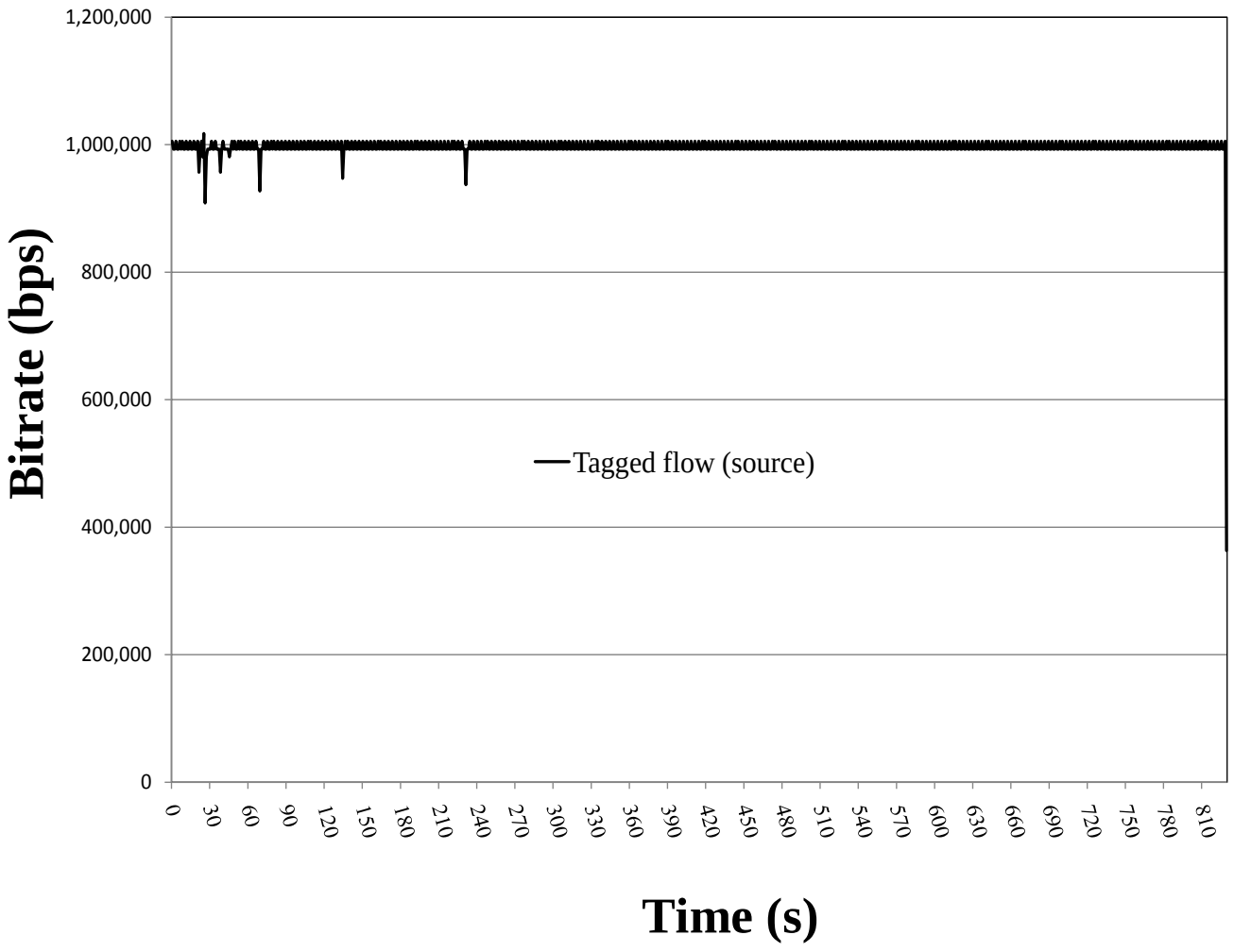

Fig.5. Tagged 1 Mbps flow observed at the ingress router a using NetMon packet capture mechanism (scenario 1)

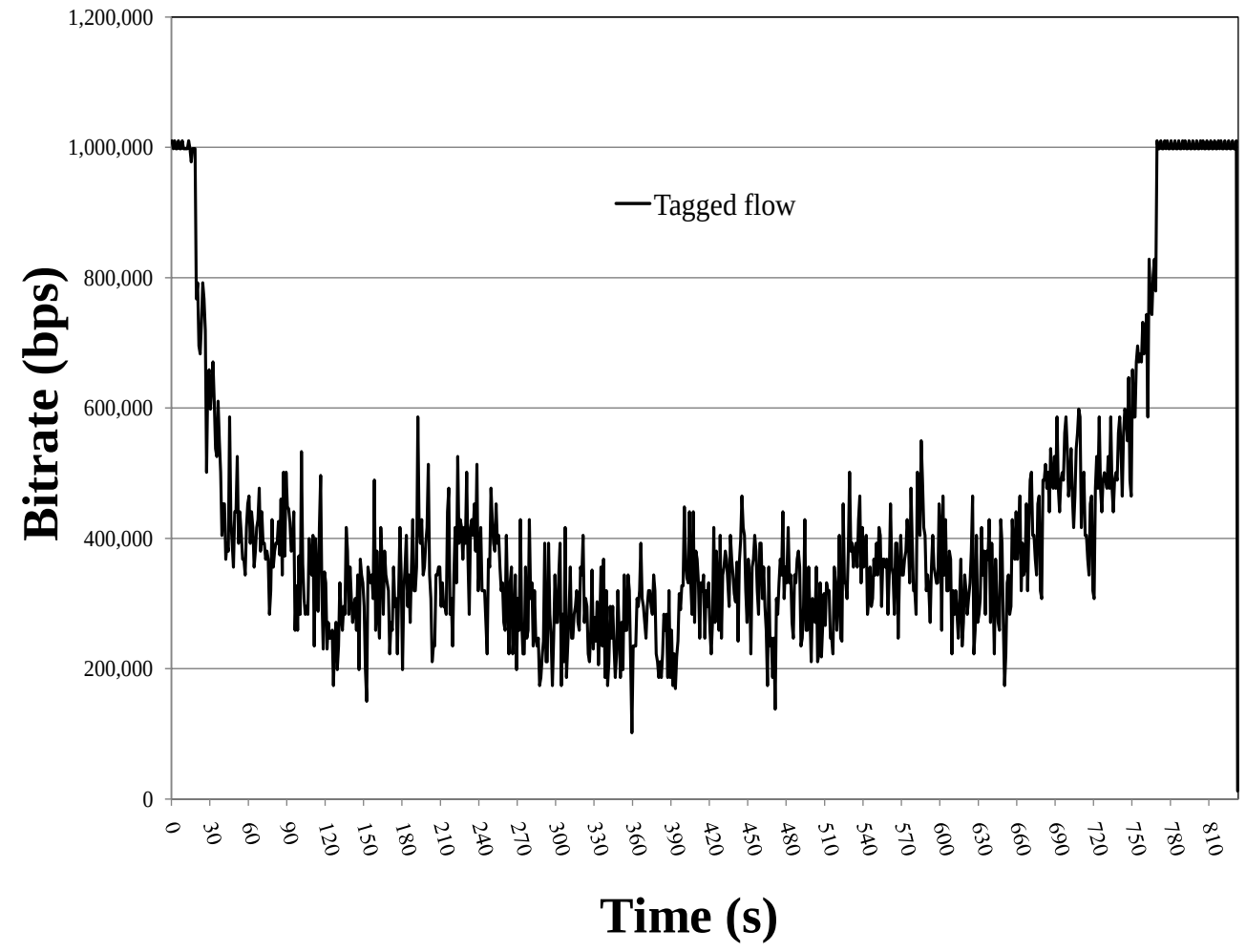

Fig.6. Tagged 1 Mbps flow observed at the egress router B. Degradation occurs due to congestion as aggregate traffic causes edge-to-edge congestion (scenario 1)

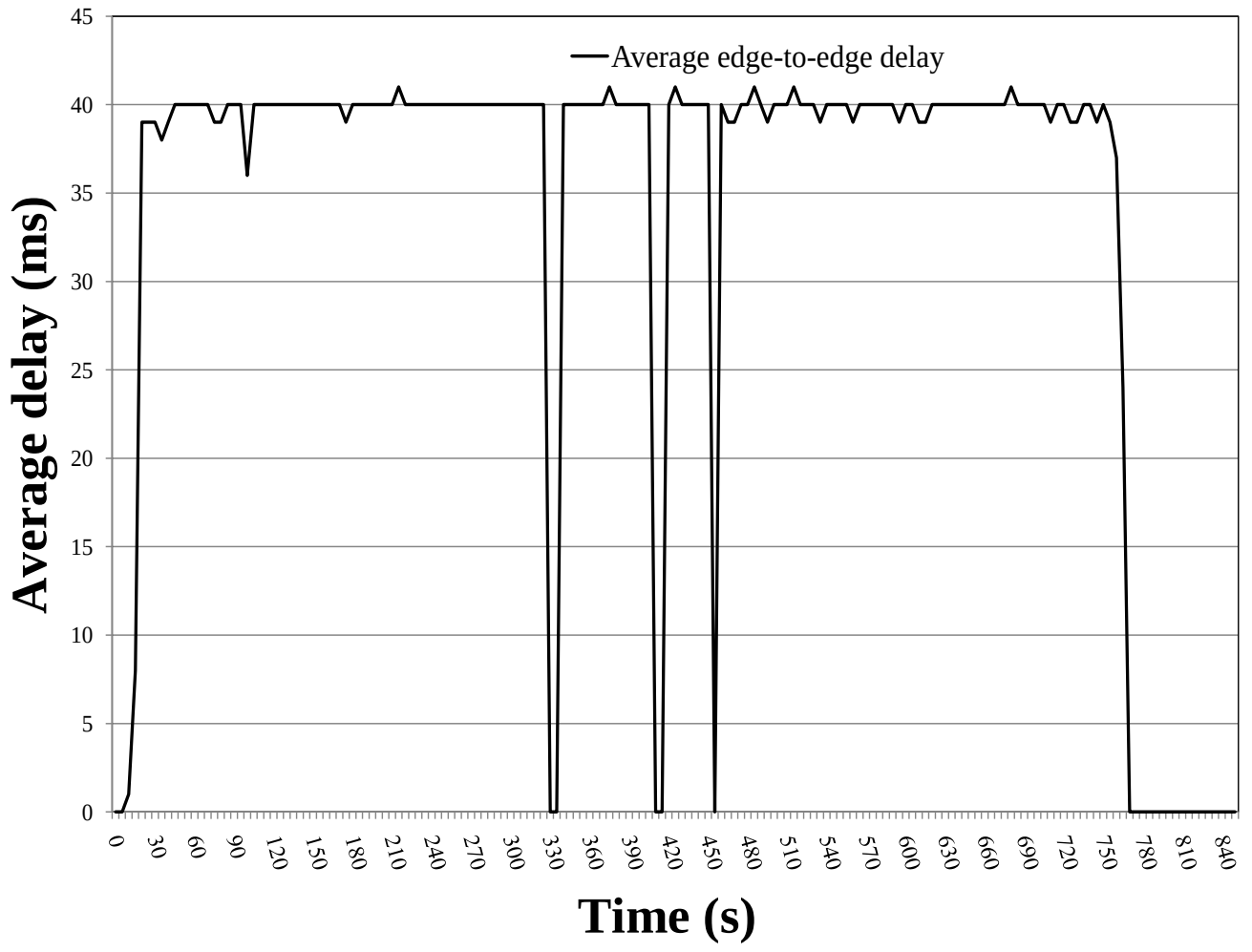

Fig.7. Average end-to-end delay without CNQF configuration policies as observed by the CNQF NetMon active measurement tool (scenario 1)

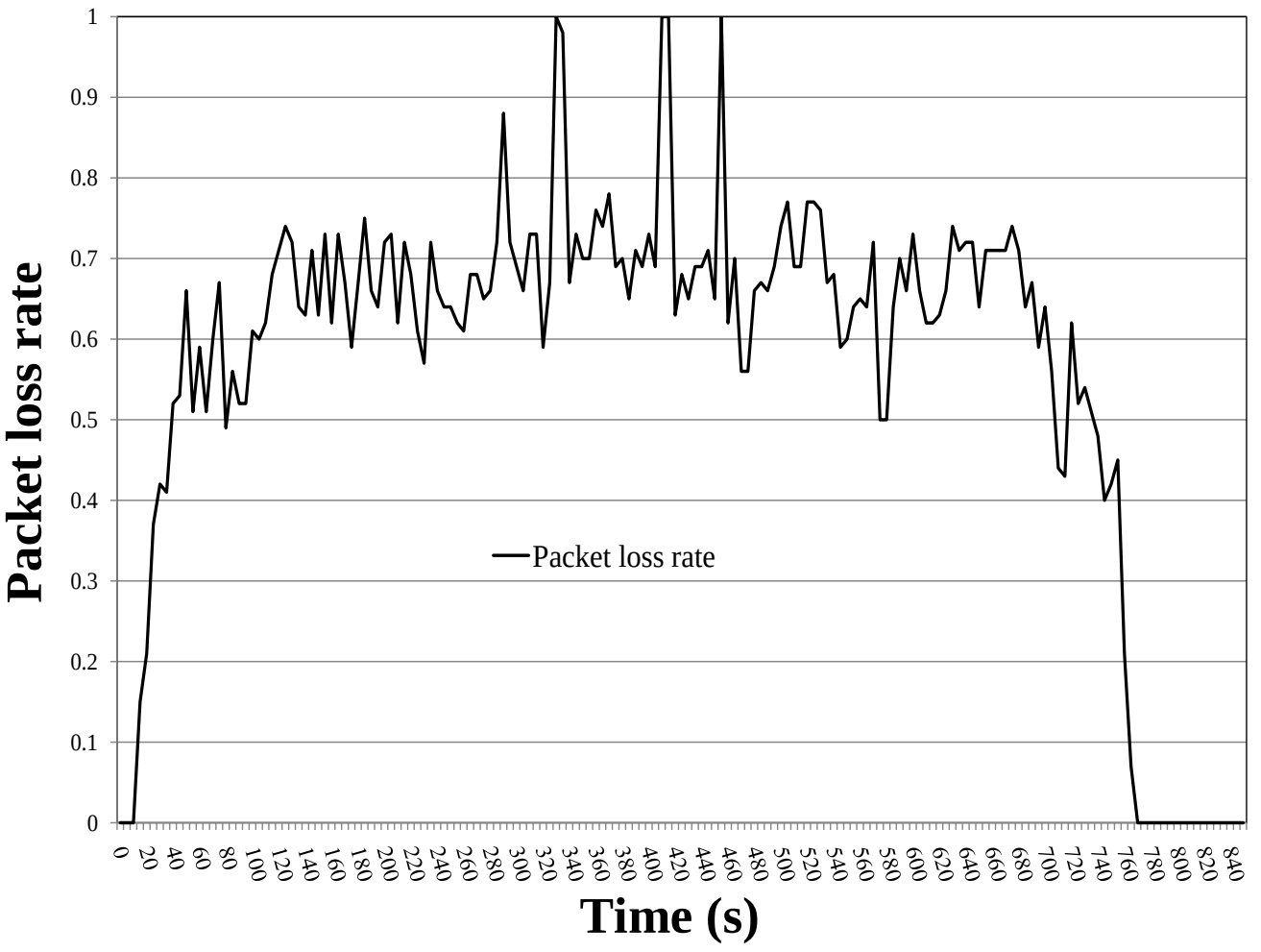

Fig.8. Packet loss rate without CNQF configuration policies as observed by the CNQF NetMon active measurement tool (scenario 1)

Fig.7 shows the measurement of average end-to-end delay from the NetMon Bwping-based active measurement mechanism depicting up to 40 ms end-to-end average delay at the times the congestion situation occur. At the same time the packet loss rate (as seen by the NetMon Bwping-based active measurement) is depicted in Fig.8 to reach above 60% most times but peaking at 100% on a few occasions thus confirming the occurrence of a congestion situation as a (result of the 100 Mbps egress bottleneck) between these times. Note that all the measurements depicted in Figs.4-9 were taken simultaneously.

### *5.2.1 Validation Tests With CNQF Configuration:*

Using the previous scenario as baseline, another test was conducted after having configured the testbed nodes (Policy Enforcement Points) with the CNQF policies as outlined in section 5.1. The aim was to validate the operation of the ResConImpl and RB implementations as well as the inter-operational functionality of the entities with the MMS NetMon as crucial building blocks of the CNQF framework.

Figs.9-13 illustrate NetMon measurements taken simultaneously while conducting the test of the second scenario with CNQF enabled via the GUI tool in running on the management station (see Fig.3). Fig.9 again shows a similar aggregate background traffic pattern to the previous scenario (Fig.4) as captured from the ingress router A. At the same time the 1Mbps flow under observation at the egress router B shows much better resilience to congestion as seen from Fig.11 compared to the previous scenario without CNQF control which resulted in the degradation observed in Fig.6.

Figs.12 and 13 show the average end-to-end delay and the packet loss rate as obtained from the NetMon Bwping-based tool. This time the measurements are configured to estimate the metrics for the priority traffic (i.e. EF traffic) by marking the probe packets also with EF DSCP within its IP header. It can be seen that a consistent pattern of average end-to-end delay falling between 28 and 29 ms is observed compared to the values of around 40 ms observed without CNQF policies (Fig.7). Likewise Fig.13 depicts a much lower loss rate recorded compared to the situation in Fig.8.

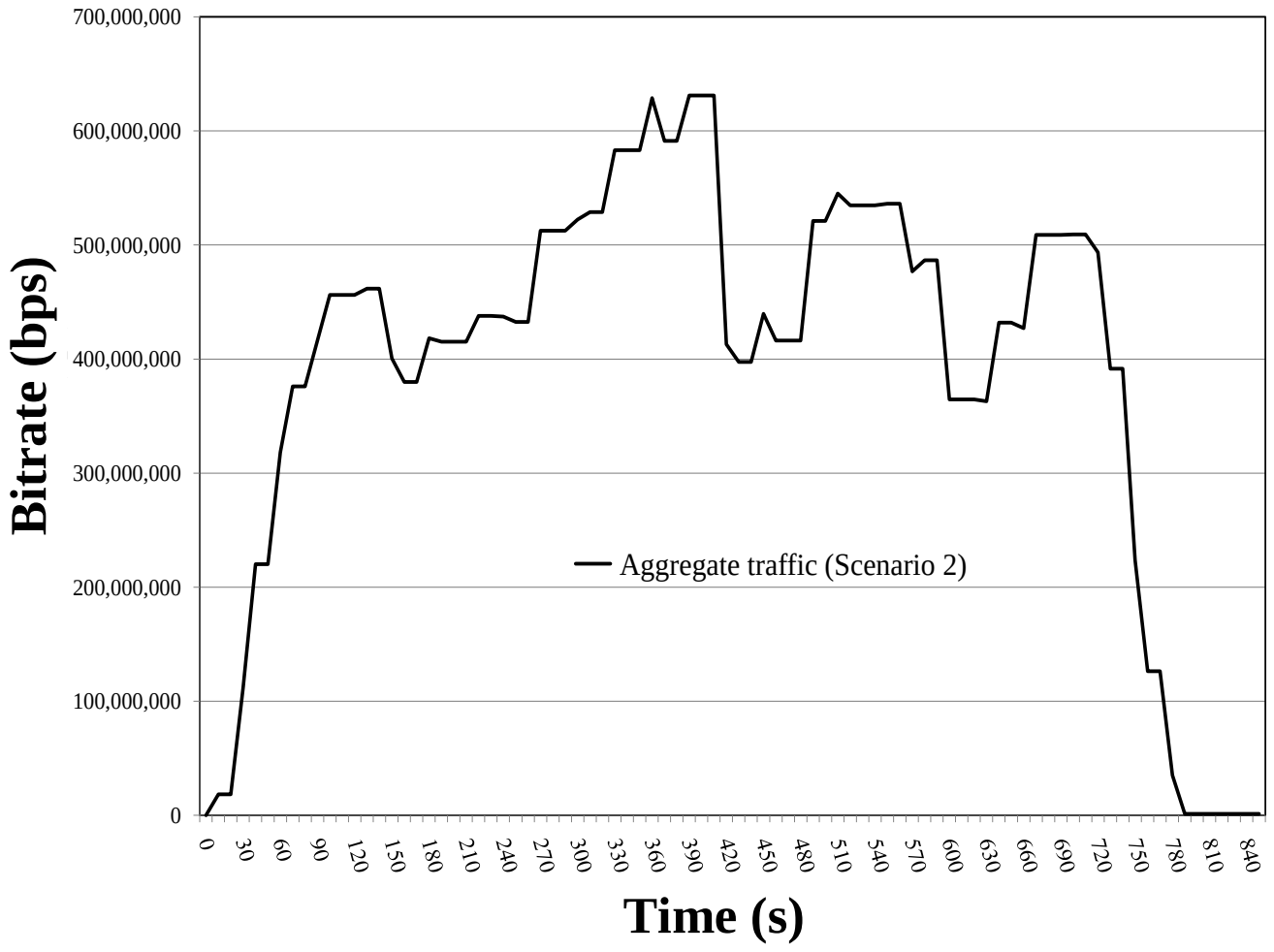

Fig.9. Aggregate traffic from all edge networks to the testbed ingress router observed from the CNQF NetMon tool (scenario2)

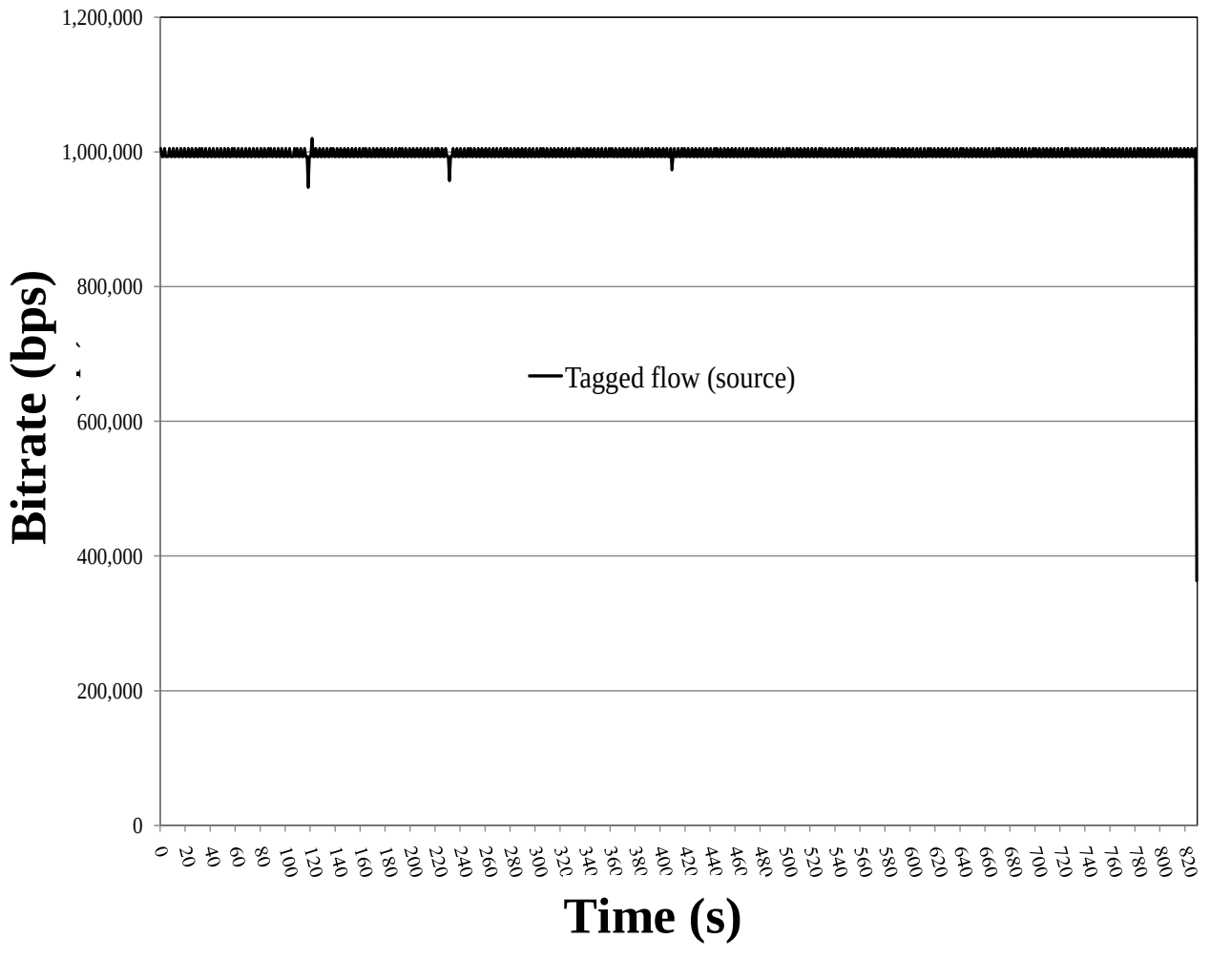

Fig.10. 1 Mbps flow observed at the ingress router A using the NetMon packet capture mechanism (scenario 2)

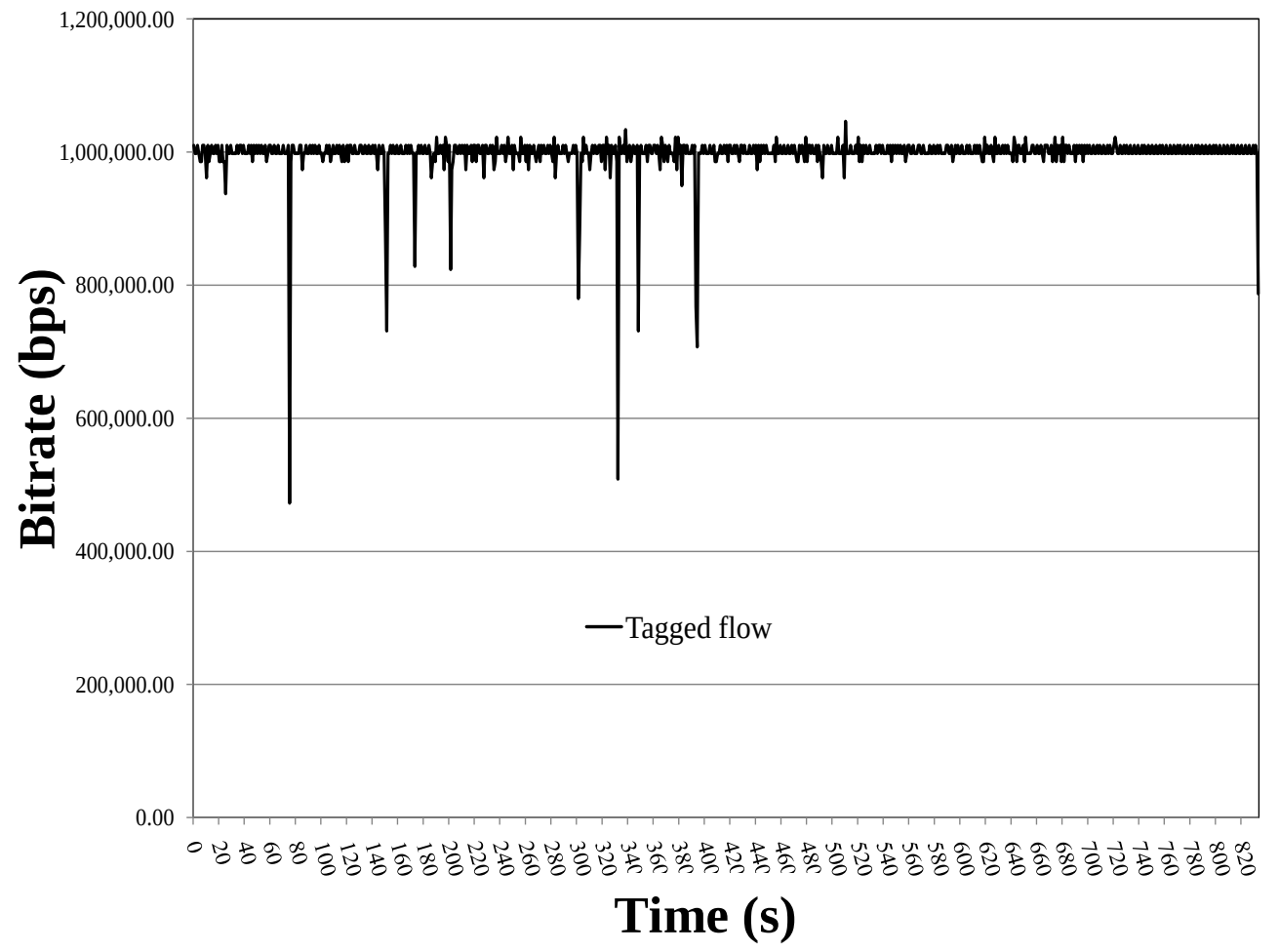

Fig.11. 1 Mbps tagged flow observed at the egress router B. Degradation is mitigated by CNQF policies even though there is congestion as aggregate background traffic changes with the pattern shown in Fig.9 (scenario 2)

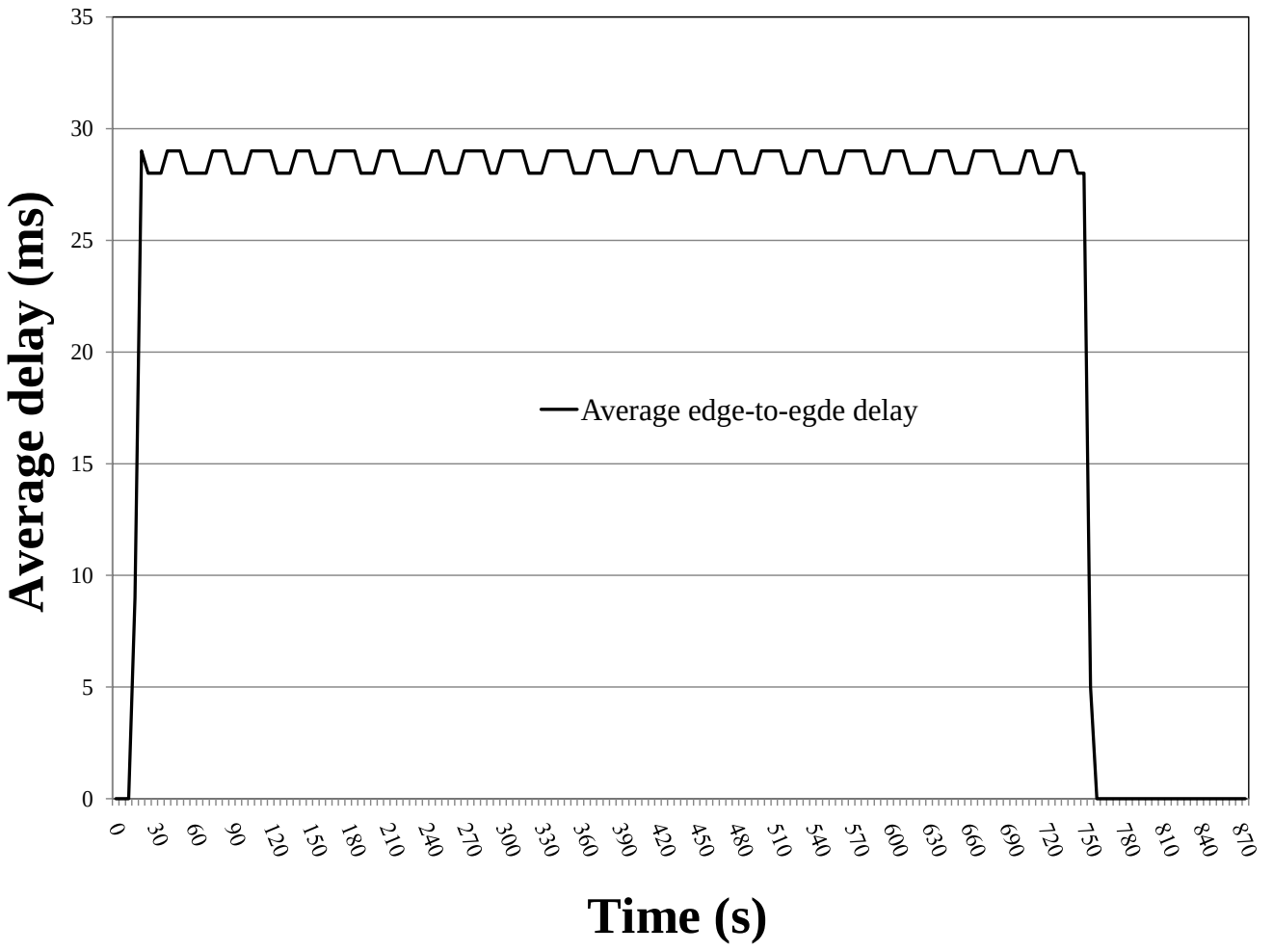

Fig.12. Average end-to-end delay with CNQF configuration policies as observed by the CNQF NetMon active measurement tool configured to measure EF traffic metrics (scenario 2)

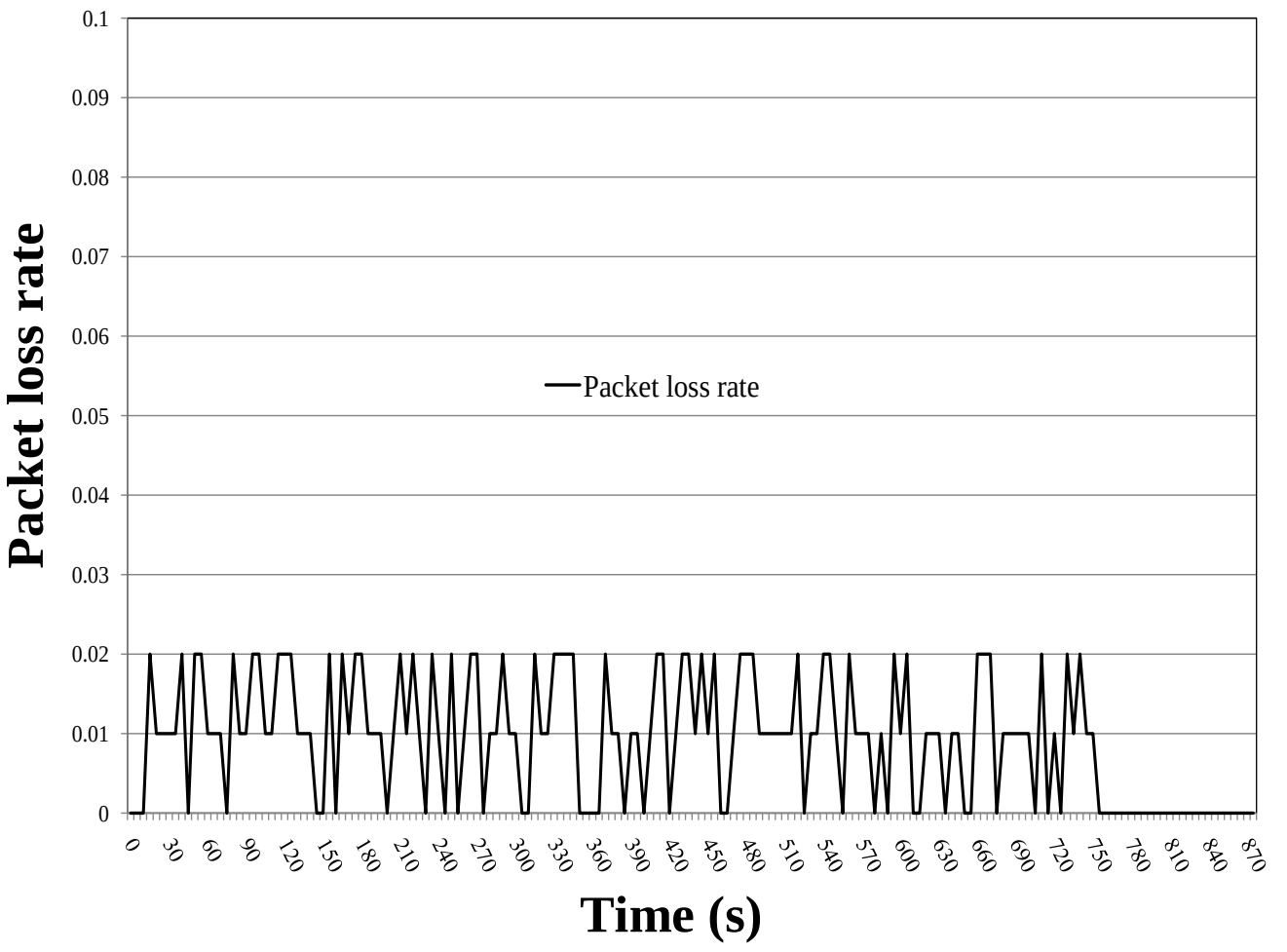

Fig.13. Packet loss rate with CNQF configuration policies as observed by the CNQF NetMon active measurement tool configured to measure EF traffic metrics (scenario 2)

The results presented indicate the successful execution of the high-level RB policies operating within the CNQF management station. This in turn validates the correct operation of the interface mechanism implemented for RB to RC communication within the framework prototype. Those in Fig.s 10 to 13 are indicative of the successful policy-driven reconfiguration of the PEPs (i.e. Linux based routers) via the implemented RC elements of the prototype. Lastly, all the measurements depicted were captured by the implemented NetMon mechanisms. Hence, from our current investigation we can conclude that CNQF RMS and MMS elements implemented within the framework prototype as well as their interoperability have been validated to be operating as expected and in accordance with the policy framework architecture.

## 6. SUMMARY AND FURTHER WORK

CNQF is designed to provide an infrastructure for policy-based management of converged networks through the various functional elements that make up its subsystems. Its design allows for closed-loop, scalable, and adaptive end-to-end QoS control in converged networks. In this paper, we presented the implementation of a Java based prototype and evaluated its ability to facilitate policy-based QoS management when deployed on a Linux-based testbed. The conducted tests validate the framework architecture as key functional elements implemented within the prototype interworked as expected. The CNQF test bed is currently under expansion to interconnect with the IU-ATC Theme 10 Mobile Network testbed located at University of Surrey and IIT Madras. When accomplished, it will allow CNQF to be deployed under a wider range of experimental scenarios.

## ACKNOWLEDGMENT

This work is funded by the EPSRC-DST India-U.K. Advanced Technology Centre of Excellence in Next Generation Networks, Systems and Services (IU-ATC). www.iu-atc.com.